\documentclass[twocolumn]{openjournal}

\usepackage{orcidlink} 
\usepackage[T1]{fontenc}
\usepackage{ae,aecompl}

\usepackage{amsmath}	
\usepackage{amssymb}	

\usepackage{natbib}

\usepackage{hyperref}
\hypersetup{
	colorlinks=true,
	urlcolor=blue,    
	linkcolor=black,  
	citecolor=blue,   
}

\hypersetup{citecolor=blue, 
            linkcolor=red, 
            menucolor=blue, 
            urlcolor=blue}  

\newcommand{\cm}{{~\rm cm}}

\newcommand{\km}{{~\rm km}}
\newcommand{\s}{{~\rm s}}

\newcommand{\g}{{~\rm g}}

\newcommand{\erg}{{~\rm erg}}
\newcommand{\yr}{{~\rm yr}}

\newcommand{\pc}{{~\rm pc}}

\begin{document}

\title{On the progenitor of the type Ia supernova remnant 0509-67.5}



\author{Noam Soker\,\orcidlink{0000-0003-0375-8987}} 

\affiliation{Department of Physics, Technion - Israel Institute of Technology, Haifa, 3200003, Israel; soker@physics.technion.ac.il}


\begin{abstract}
Based on the iron and hydrogen similar elliptical morphologies of the type Ia supernova (SN Ia) remnant (SNR)~0509-67.5, I suggest that the ambient gas shaped the SN ejecta, that it is a remnant of an old planetary nebula, and that the explosion was spherical. 
Adding that there is no observed stellar survivor in this SNR, known to be an SN Ia, I conclude that the SN Ia scenarios that best account for the explosion of this SN inside a planetary nebula (SNIP) are the lonely white dwarf scenarios,  i.e., the core-degenerate (CD) and the double degenerate (DD) with merger to explosion delay (MED) time, i.e., the DD-MED scenario. Other scenarios encounter challenges, but I cannot completely rule them out. If true, the suggestion of SNR~0509-67.5 being an SNIP implies that the fraction of SNIPs might be higher than the previously estimated 50 percent of all normal SNe Ia. In the frame of the CD and the DD-MED scenarios, 
I attribute the observed double-shell structure in calcium to Rayleigh-Taylor instability during the explosion process: the caps of the Rayleigh-Taylor instability mushrooms form the outer shell. Instabilities during the explosion process might form clumps of some elements moving much faster than their mean velocity, possibly explaining fast-moving intermediate-mass elements.
\end{abstract}


\section{INTRODUCTION}
\label{sec:intro}

Several theoretical scenarios exist for type Ia supernovae (SNe Ia), some with two or more channels. Because there is no agreement in the community on the classification, I cannot state the number of scenarios and channels. The relatively large number of reviews in the last decade reflects the disagreement on the dominant scenarios of SNe Ia and the way to classify them (\citealt{Maozetal2014, MaedaTerada2016, Hoeflich2017, LivioMazzali2018, Soker2018Rev, Soker2019Rev, Soker2024Rev, Wang2018,  Jhaetal2019NatAs, RuizLapuente2019, Ruiter2020, Aleoetal2023, Liuetal2023Rev, Vinkoetal2023, RuiterSeitenzahl2025}).
All SN Ia scenarios have some advantages over the others. On the other hand, all SN Ia scenarios encounter challenges in explaining some observations, overcoming theoretical difficulties, or both (e.g., \citealt{Pearsonetal2024, SchinasiLembergKushnir2024, sharonKushnir2024, Sharonetal2024, Wangetal2024}). Some scenarios and channels might account for peculiar SNe Ia but not for normal SNe Ia. Given the above, studies of SN Ia scenarios that concentrate on one, two, or three scenarios should also consider the rest.

I present in Table \ref{Tab:Table1} the classification from \cite{Soker2024Rev} where more properties of the six scenarios of this classification are listed (see also \citealt{BraudoSoker2024, BraudoSoker2025}). All these scenarios have been actively studied in recent years (e.g., \citealt{Boraetal2024, Bregmanetal2024, DerKacyetal2024, Itoetal2024, Koetal2024, Kobashietal2024, Limetal2024, Palicioetal2024, Phillipsetal2024, Soker2024RAAPN, Uchidaetal2024, OHoraetal2025, ShenK2025}, out of many more papers just in the last two years; for most recent reviews see \citealt{Soker2024Rev} and \citealt{RuiterSeitenzahl2025}).  
\begin{table*}
\scriptsize
\begin{center}
  \caption{An SN Ia scenarios classification}
    \begin{tabular}{| p{1.8cm} | p{2.4cm} | p{2.4cm}| p{2.0cm}| p{2.0cm} | p{2.0cm} | p{2.0cm} |}
\hline  
\textbf{Group} & \multicolumn{2}{c|}{$N_{\rm exp}=1$: {{Lonely WD}}}  &  \multicolumn{4}{c|}{$N_{\rm exp}=2$}     \\  
\hline  
\textbf{{SN Ia Scenario}}  & {Core Degenerate }    & {Double Degenerate - MED} & {Double Degenerate} & {Double Detonation} & {Single Degenerate} & {WD-WD collision} \\
\hline  
\textbf{{Name}} & CD & DD-MED & DD & DDet & SD-MED or SD & WWC\\
\hline  
\textbf{MED time} & MED  & MED  & 0  &0  & MED or 0 & 0 \\
\hline  
 {$\mathbf{[N_{\rm sur}, M, Ej]}$$^{[{\rm 2}]}$}
  & $[0,M_{\rm Ch},{\rm S}]$ 
  & $[0,M_{\rm Ch}, {\rm S}]$ 
  & $[0,$sub-$M_{\rm Ch},{\rm N}]$
  & $[1,$sub-$M_{\rm Ch},{\rm N}]$
  & $[1,M_{\rm Ch},{\rm S~or~N}]$  
  & $[0,$sub-$M_{\rm Ch},{\rm N}]$ \\
\hline  
\textbf{{SNR 0509-67.5}} & Most Likely & Likely & Possible & Less possible  & Unlikely & Unlikely  \\
\hline  
     \end{tabular}
  \label{Tab:Table1}\\
\end{center}

\begin{flushleft}
\small 
Notes: An SN Ia scenarios classification scheme from \cite{Soker2024Rev}. The DDet scenario with two exploding WDs is grouped in this table with the DD scenario. The last row is of this study, with the estimated likelihood of each scenario to account for SNR~0509-67.5 
\newline
 Abbreviation. MED time: Merger to explosion delay time or mass transfer to explosion delay time.    
$N_{\rm exp}$: system's number of stars at the explosion moment. $N_{\rm sur}$: if a companion survives the explosion, then $N_{\rm sur}=1$, while if no star survives in the system $N_{\rm sur}=0$. In some peculiar SNe Ia, the exploding WD survives, and the system can have $N_{\rm sur}=2$. $M_{\rm Ch}$ and sub-$M_{\rm Ch}$ mark a near-Chandrasekhar-mass and sub-Chandrasekhar mass explosions, respectively. Ej is the ejecta morphology: S indicates scenarios that can lead to a spherical SNR, while N indicates scenarios that expect to form SNRs with large departures from sphericity. 
\end{flushleft}
\end{table*}

This study focuses on SNR~0509-67.5 (J0509-6731), a type Ia SN remnant (SNR) in the Large Magellanic Cloud (LMC) that attracted many studies (e.g., \citealt{Smithetal1991, WarrenHughes2004, Borkowskietal2006, Ghavamianetal2007, Kosenkoetal2008, Seoketal2008, Williamsetal2011, Bozzettoetal2014, Katsudaetal2015, Kosenkoetal2015} and others cited in this study). In Section \ref{sec:Properties}, I present the relevant properties of SNR~0509-67.5. 
 This study accepts the common view that SNR~0509-67.5 is an SN Ia and focuses on the question of the progenitor scenario and the presence of a circumstellar matter (CSM). 

In a recent study, \cite{Dasetal2024Poster} explained the double-shell segments of SNR~0509-67.5 with the double-detonation (DDet) scenario (Section \ref{subsubsec:DDet} here). In the DDet scenario, the pre-explosion system is a binary system with a CO white dwarf (WD) accreting helium from a helium star or another WD. The ignition of the helium layer on the surface of the mass-accreting WD excites a shock wave that propagates inwards and detonates the WD. Many groups in recent years studied the DDet scenario (e.g., just from 2024, \citealt{Callanetal2024, MoranFraileetal2024, PadillaGonzalezetal2024, Polinetal2024, Shenetal2024, Zingaleetal2024, Glanzetal2024, Rajaveletal2025}).  
The DDet has a channel where both the mass-accretor WD and the mass-donor WD explode, leaving no survivor; in Table \ref{Tab:Table1}, this channel is grouped with the DD scenario.
There is the triple-detonation channel where a helium WD mass-donor explodes (e.g., \citealt{Papishetal2015, Bossetal2024}), and its quadruple-detonation sub-channel where the outer helium layer of the mass-donor HeCO WD explodes and detonates the CO interior part. 
 (e.g., \citealt{Tanikawaetal2019, Pakmoretal2022}). \cite{Pakmoretal2021} simulate a case where the explosion of the helium layer on the mass-accretor WD fails to detonate the CO inner part of the WD but detonates the mass-donor WD. This leaves the mass-accretor WD as the surviving companion. 

In Section \ref{sec:Progenitor}, I propose alternative explanations for the double-shell segments that do not need the DDet scenario. I summarize this study in Section \ref{sec:Summary}, strengthening the call to keep all scenarios in mind when studying SNe Ia.

\section{SNR 0509-67.5 might be a supernova inside a planetary nebula (SNIP)}
\label{sec:Properties}

Many studies have shown that the east and west sides of SNR~0509-67.5 are unequal in several properties, like texture (in the west, there are thin H$\alpha$ filaments; e.g., \citealt{Litkeetal2017, Hoveyetal2018}), brightness (the west is brighter in many wavelengths), and shock properties (e.g., \citealt{Helderetal2010}).

\begin{figure}[t]
\includegraphics[trim=0.0cm 8.5cm 15.0cm 0.0cm,scale=0.81]{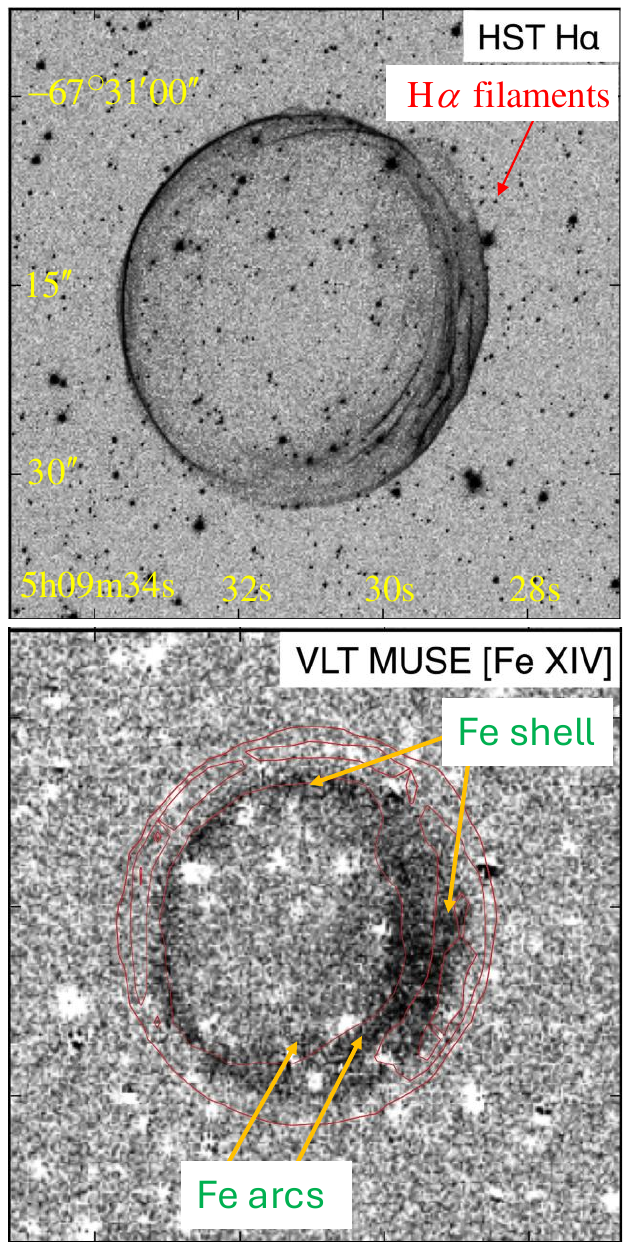} 
\caption{Images adapted from \cite{LiChuRaymondetal2021}. (a) An HST H$\alpha$ image. Vertical and horizontal axes are declination and right ascension (J2000). I mark the H$\alpha$ filaments.    
(b) VLT MUSE [Fe XIV] image with VLT MUSE H$\alpha$ contours in red. Both panels have the same scale. I added the marks of the Fe shell and arcs.  
}
\label{Fig:Halpha}
\end{figure}

In Figure \ref{Fig:Halpha} I present H$\alpha$ and iron images of SNR~0509-67.5 adapted from \cite{LiChuRaymondetal2021}. This figure reveals (1) that the hydrogen is outside the iron, (2) the H$\alpha$ filaments on the west and south of the SNR, and (3) the Fe shell and two arcs. The shell is an iron-bright outer ellipse that is closed, and the arcs are bright, narrow zones extending from the south to the north. The arcs might be filaments or the projections of caps, but they are not closed shells. 

\begin{figure*}[t]
\begin{center}
\includegraphics[trim=0.0cm 10.5cm 0.0cm 0.0cm,scale=0.81]{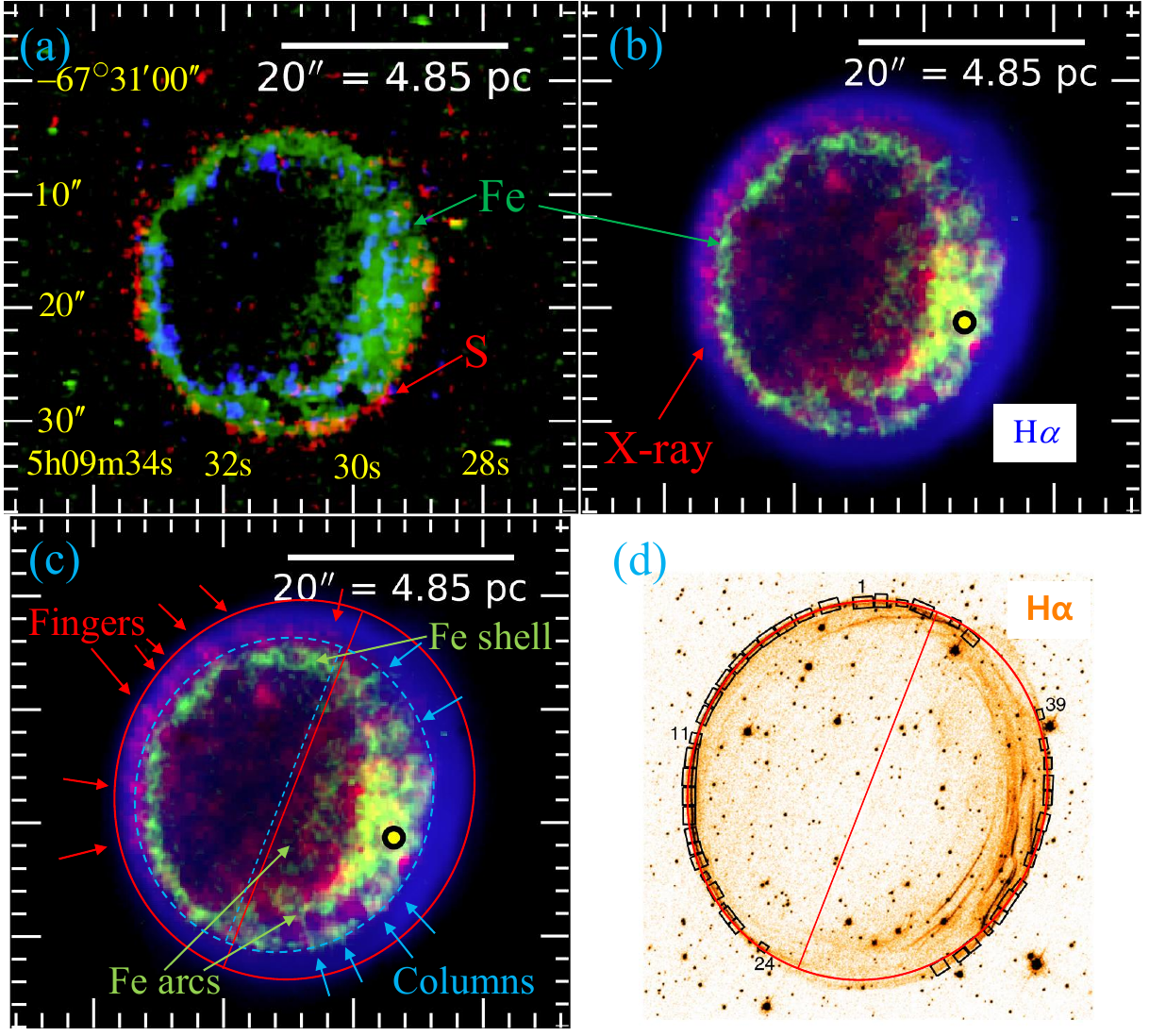} 
\caption{ (a-c): Images of SNR~0509-67.5 adapted from \cite{Seitenzahletal2019}. (a) A composite image: red: [S \textsc{xii}] $7613.1 \text{\AA}$; blue: [Fe \textsc{ix}] $8236.8 \text{\AA}$; green: [Fe \textsc{xiv}] $5303 \text{\AA}$. (b) A composite image of X-rays from Chandra ACIS in red (for an X-ray image, see also, e.g., \citealt{WarrenHughes2004}), H$\alpha$ in blue (VLT-MUSE), and [Fe \textsc{xiv}] in green (VLT-MUSE). (Yellow dot: region of spectra extraction.) (c) It's like panel b but with other marks that I added. The red solid-line ellipse is as in panel d. The pale-blue dashed-line ellipse is 0.83 times that of the red ellipse and has the same orientation. It is displaced to mark the edge of the [Fe \textsc{xiv}] zone. I attribute the X-ray fingers and Fe columns to Rayleigh-Taylor instabilities. (d) An H$\alpha$ image by ACS-HST adapted from \cite{Hoveyetal2015}. I added an ellipse through the rectangles that \cite{Hoveyetal2015} drew on the exterior H$\alpha$ rim. The straight red line is the ellipse's long axis, which is 1.075 times as long as the short axis. } 
\label{Fig:Fe}
\end{center}
\end{figure*}

Figure \ref{Fig:Fe} presents three more images: iron and sulfur (panel a), iron emission, X-ray, and hydrogen (panel b and c), and hydrogen with the bright rim marks by many rectangles that \cite{Hoveyetal2015} used to calculate the SNR expansion. The bright rim that \cite{Hoveyetal2015} mark forms an ellipse, as indicated by the solid red line on panel d; the straight red line is the ellipse's long axis. I copied this ellipse to panel c, which falls on the outer boundary of the H$\alpha$ emission in blue. I reduced the size of the ellipse by a factor of 0.83 and displaced it to the southeast to match the boundary of the iron emission; the reduced-displaced ellipse fits the iron boundary. The X-ray map shows `fingers' on the west and north (pointed at by red arrows). The iron map shows radial extensions from the outer Fe-arc to the Fe shell (pointed at by pale-blue arrows). I attribute the fingers and columns to Rayleigh-Taylor instabilities that I discuss further in Section \ref{sec:Progenitor}.  As Rayleigh-Taylor instabilities occur in many types of flow, the formation of fingers, mushrooms-shaped protrusions, and clumps is very common, e.g., they can be observed in core-collapse supernova remnants (e.g., \citealt{Alsaberietal2024}).

\cite{Guestetal2022} study the expansion of SNR~0509-67.5. They find the average expansion velocity to be $6120 \km \s^{-1}$ with large variations $4900 - 7360 \km \s^{-1}$.
\cite{Arunachalametal2022} measure the expansion velocity to be $6315 \pm 310 \km \s^{-1}$ and estimate the SNR age as $315.5 \pm 1.8 \yr$ with ambient medium densities range of $3.7 - 8.0 \times 10^{-25} \g \cm^{-3}$. 
The shock radius is $R_s= 3.66 \pm 0.036 \pc$. 
For a constant pre-explosion ambient density of $\rho_a=6 \times 10^{-25} \g \cm^{-3}$ the swept-up mass is $M_{\rm su} \simeq 1.8 M_\odot$. 
 
Based on the expansion of SNR~0509-67.5, \cite{Arunachalametal2022} calculate the pre-explosion ambient density near the rim of the remnant and find two peaks in the east, position angle $95^\circ$, and west-south-west, $245^\circ$; the density minima are in the north, $\simeq 0^\circ$, and south, $170^\circ$. 
These maxima and minima are in directions within $\simeq 15^\circ$ from the short and long axes of the ellipse drawn in Figure \ref{Fig:Fe}.  
The similar elliptical shape of the hydrogen-rich ambient gas and the iron-rich ejecta suggests that either the ejecta shaped the ambient gas or vice versa. The higher ambient gas density along the ellipse's short axis shows that the ambient gas shaped the ejecta. The brighter emission in the west-south-west side, e.g., in iron lines and X-ray, probably results from a much stronger reverse shock in that direction due to dense ambient gas there, in particular in front of the ejecta or behind it and to the west, namely, the part projected near the outer iron arc.  

The ambient swept mass I estimated above is $M_{\rm su} \simeq 1.8 M_\odot$; with more mass outside the remnant, this adds up to a total ambient mass of $M_{\rm am} > 1.8 M_\odot$. The elliptical morphology of the hydrogen-rich ambient medium and its mass are compatible with the ambient gas being a remnant of a very old planetary nebula, an age of $>10^5 \yr$. Namely, it is possible that SNR~0509-67.5 exploded inside a remnant of an elliptical planetary nebula, i.e., an SNIP (SN inside a planetary nebula).  The planetary nebula should have been elliptical because it shaped the outer SNR 0509-67.5, which is elliptical. Based on the mass and density alone,  I cannot rule out the fact that the ambient gas is an interstellar medium (ISM). However, the elliptical morphology strongly suggests an SNIP because it is highly unlikely for an ISM to have this morphology centered on the explosion. 
If SNR~0509-67.5 is an SNIP, it could come from the CD, the DD-MED, the DD, or the DDet scenarios. In the CD scenario, the core-WD merger occurs during the common envelope evolution (CEE). In contrast, in the DD-MED, DD, and DDet scenarios, the merger occurs much later, likely due to gravitational wave emission by the core remnant (the younger WD) and the older WD binary system.  
The CEE to explosion delay (CEED; see \citealt{Soker2022RAA}) time is 
$t_{\rm CEED} = 3.66 \pc / v_{\rm am} = 3.6 \times 10^5  (v_{\rm am}/10 \km \s^{-1})^{-1} \yr$, where $v_{\rm am}$ is the expansion velocity of the ambient medium at the rim of the remnant (there is more ambient gas outside the rim), which in this model is the former common envelope. 
The CEED time, if SNR~0509-67.5 is an SNIP, is long.

From the study of the ejecta interaction with the planetary nebula, \cite{Courtetal2024} argue that the CEED time of SNe Ia should be $t_{\rm CEED} > 10^4 \yr$ (however, some SNe Ia have interaction with much closer CSM). The CEED time of  SNR~0509-67.5 obeys their requirement.  
 
 \cite{Yamaguchietal2014} place SNe (core-collapse supernovae and SNe Ia) on a plane of Fe K$\alpha$ luminosity versus its centroid energy. They find that the two groups occupy different areas in that plane. Indeed, SNR~0509-67.5 location is far from the area of core-collapse supernovae, compatible with its known classification as an SN Ia. \cite{Yamaguchietal2014} explain this separation of supernova types on this diagram as resulting from core-collapse supernovae interacting with a dense CSM, while SNe Ia not. However, I note that the Kepler SNR is known to interact with a dense CSM (e.g., \citealt{Patnaudeetal2012}), but it is still far from the area of core-collapse supernovae in this plane. Therefore, I argue that the location of SNR~0509-67.5 close to Kepler SNR in that plane might show that the location of SNR~0509-67.5 on that plane does not rule out an interaction with a CSM.  

Studies in recent years have argued that more SNRs Ia should be SNIPs. Type Ia SNR0519-69.0 in the LMC is an example. 
\cite{TsebrenkoSoker2015a} classified it as a `maybe SNIP', \cite{Soker2022RAA} classified it as an SNIP based on the presence of dense CSM, and \cite{Schindelheimetal2024} solidify the SNIP classification of SNR~0519-69.0 by quantitative calculation of the ejecta-CSM interaction and its X-ray spectral. In my previous classification in \cite{Soker2022RAA}, I list SNR~0509-67.5 as a non-SNIP. I have now changed this to \textit{maybe SNIP}. 

\section{The progenitor of SNR 0509-67.5}
\label{sec:Progenitor}
\subsection{Key properties}
\label{subsec:KeyProperties}

I consider the following properties to examine the likelihood of an SN Ia scenario to account for SNR~0509-67.5. 
 
\textit{(1) No stellar remnant.} 
Searches (e.g., \citealt{SchaeferPagnotta2012, Panetal2014, Notaetal2016, Shieldsetal2023}) find no surviving companion in SNR~0509-67.5. This limits the parameter space for the SD scenario and the DDet with a surviving companion scenario, allowing only an M dwarf mass-donor companion in the single degenerate scenario, as \cite{Wheeler2012} proposed.
\cite{DiStefanoKilic2012} and \cite{MengPodsiadlowski2013} consider the scenario where the WD needs to lose angular momentum before it explodes (\citealt{DiStefanoetal2011, Justham2011}) and argue that one should be more cautious with the claim of no surviving companion in the frame of the SD scenario.  
 
\textit{(2) Energetic explosion.}
Based on light-echo from SNR~0509-67.5 \citep{Restetal2005}, \cite{Badenesetal2008} and \cite{Restetal2008} argue that SNR~0509-67.5 was a 1991T-like SN Ia, implying a bright SN Ia. \cite{Badenesetal2008} estimate the explosion energy to be $1.4 \times 10^{51} \erg$ and nickel-56 production of $0.97 M_\odot$. This high mass of nickel-56 makes it unlikely that a single sub-Chandrasekhar-mass WD exploded. It can be a (close to) Chandrasekhar mass explosion or two sub-Chandrasekhar mass WDs exploded.

I also give a high weight to my tentative claims in Section \ref{sec:Properties}: 
\newline
\textit{(3) SNR~0509-67.5 is an SNIP.} Namely, the ambient medium of SNR~0509-67.5 is an old remnant of a planetary nebula (which swept some ISM), a planetary nebula age of $t_{\rm CEED} \simeq 10^5 - 5 \times 10^5 \yr$. 
\newline
\textit{(4) Spherical, or close to, explosion. }
The ambient medium, a former planetary nebula, shaped the SN ejecta. Namely, the explosion itself was spherical or slightly different from a spherical one. Had the explosion been highly-non-spherical, the SNR morphology  would largely depart elliptical one.

\subsection{The possible scenarios for SNR~0509-67.5}
\label{subsec:Scenarios}

I estimate the likelihood of the different scenarios to explain SNR~0509-67.5, as summarized in the last row of Table \ref{Tab:Table1}. 

\subsubsection{The lonely WD scenarios: CD and DD-MED}
\label{subsubsec:CDDDMED}
The CD scenario predicts that many SNe Ia are SNIPs (e.g., \citealt{TsebrenkoSoker2015a, Soker2022RAA}; \citealt{Soker2024Rev} for a review). It also predicts a spherical,  or rarely, elliptical explosion if the WD rapidly rotates.  I consider this scenario the most likely to explain the properties of SNR~0509-67.5.  

In the DD-MED, the merger product of the two WDs explodes only after it relaxes dynamically, leading to a spherical (or elliptical explosion if rapidly rotating).  If the merger and explosion take place within $t_{\rm CEED} \simeq 10^5 - 5 \times 10^5 \yr$ (Section \ref{subsec:KeyProperties}), it leads to a similar outcome to the CD scenario. I consider it less likely than the CD scenario because the gravitational waves that cause the two WDs to merge should cause these two WDs (an old WD and the remnant of the asymptotic giant branch star) to merge within the time  $t_{\rm CEED}$. This, in turn, requires the old WD to end the common envelope evolution very close to the core of the asymptotic giant branch star. 

\cite{Woodsetal2018} find that the hydrogen around SNR~0509-67.5 is mostly neutral. The CD scenario explains this as an old planetary nebula whose central star had already cooled and had low luminosity at the explosion, so the nebula had time to recombine. In the DD-MED scenario, either the merger occurred shortly after the common envelope revolution and the merger product had time to cool or shortly before the explosion, such that the hot merger remnant had no time to ionize the planetary nebula.

In the CD scenario, the second star to evolve, the one that, when it becomes an asymptotic giant branch star, engulfs the WD, forces the WD to merge with the core. The envelope should be massive enough, i.e., $M_{\rm env} \gtrsim 3 M_\odot$, implying an asymptotic giant branch star of $\gtrsim 3.5 M_\odot$. As the secondary star can accrete mass from the primary, its mass on the zero-age main sequence (ZAMS) can be lower than this value. The condition is $M_{2,{\rm ZAMS}} \gtrsim 2 M_\odot$ (e.g., \citealt{IlkovSoker2013}), with large uncertainties due to the poorly determined common envelope evolution outcomes. Due to large uncertainties, I also consider a more stringent constraint of  $M_{2,{\rm ZAMS}} \gtrsim 3 M_\odot$. 
The mass limit implies that in the SNIP channel of the CD scenario, the SNR comes from a population of an age of $\tau_{\rm SF} \lesssim 5 \times 10^8 \yr -  1.5 \times 10^9 \yr$. \cite{Badenesetal2009} argue that SNR~0509-67.5 resides in a population with a mean age of $\tau_{\rm SF} \simeq 7.9 \times 10^9 \yr$. 
However, they still leave it possible that a fraction of this population has an age of $\lesssim 1.8 \times 10^8 \yr$, i.e., progenitors with $M_{2,{\rm ZAMS}} \simeq 4-6 M_\odot$. I suggest here that SNR~0509--67.5 comes from that population.   

\subsubsection{The SD scenario}
\label{subsubsec:SD}

With the observations listed in Section \ref{subsec:KeyProperties}, I find the SD scenario unlikely. If the companion to the exploding WD supplied the ambient gas, it was an asymptotic giant branch star. If the companion avoids detection, it must presently be the cold WD remnant of the asymptotic giant branch star. This implies that the mass transfer occurred a long time ago, probably more than the age of the CSM. Although I cannot completely rule out the SD scenario for SNR~0509-67.5 with MED time, I consider it less likely than some of the other scenarios I consider below.

\subsubsection{Two exploding WD scenarios: DD and WWC}
\label{subsubsec:TwoWDs}

The WWC scenario, where two unbound WDs collide and both explode, is unlikely to account for SNR~0509-67.5 for two reasons. Firstly, it produces a highly non-spherical explosion (e.g., \citealt{Kushniretal2013, Glanzetal2023}). Secondly, it is not expected to have an elliptical ambient gas.

The two WDs exploded in the DD scenario, which in this study classification includes the DD+DDet scenario, where a helium outer layer detonation does the explosion of at least one WD. As in the WWC scenario, the explosion of both WDs leads to a highly non-spherical explosion (e.g., \citealt{Tanikawaetal2019, Polinetal2024}).
To account for the ambient gas, if it comes from the progenitor, as I claim in this study, the WD interaction should be within the time $t_{\rm CEED} \simeq 10^5 - 5 \times 10^5 \yr$. This is possible but rare. I consider the DD (and DD+DDet) scenarios possible but not as likely as the CD and DD-MED scenarios.   

\subsubsection{The DDet scenario}
\label{subsubsec:DDet}

The detonation of the helium outer layer and then the inner CO WD in the DDet scenario form two shells  (e.g., \citealt{Collinsetal2022}). The Fe arcs on the west (Figures \ref{Fig:Halpha} and \ref{Fig:Fe}) are not compatible with shells that the DDet explosion forms in the simulations by \cite{Collinsetal2022}. In the simulations, the shells appear as full rings, closed at $360^\circ$, rather than arcs of $\simeq 180^\circ$ as the Fe arcs of SNR~0509-67.5. 

\begin{figure*}[t]
\begin{center}
\includegraphics[trim=0.0cm 15.5cm 5.0cm 0.0cm,scale=1.11]{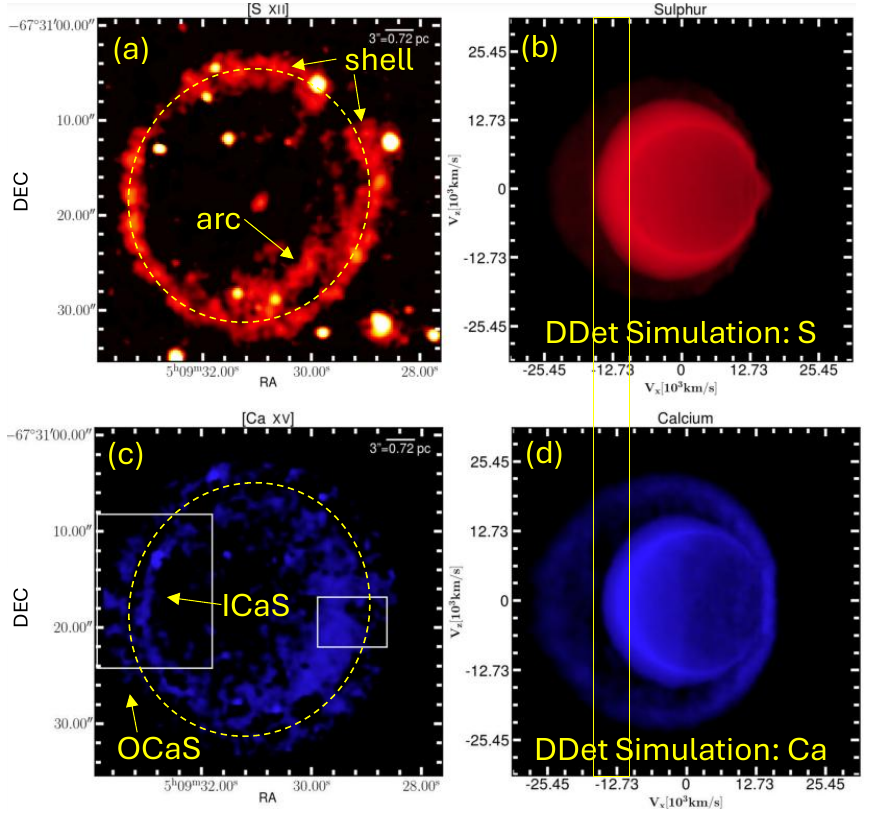} 
\caption{A figure adapted from \cite{Dasetal2024Poster}. Panels a and c are MUSE-VLT observations. Panels b and d are column densities of sulfur and calcium from DDet simulations with CO WD mass of $1 M_\odot$ and helium shell mass of $0.03 M_\odot$ from \cite{Collinsetal2022}. I added the yellow marks, including the two vertical lines extending from panel b to d, and the ellipse on panel a and c, the same on panels a and c. This ellipse has the same axes ratio and same orientation as that in Figure \ref{Fig:Fe}. It is scaled to match the sulfur shell and between the two calcium shells, showing that the observed sulfur shell is between the calcium shells \citep{Dasetal2024Poster}.   
ICaS: Inner calcium shell. OCaS: Outer calcium shell. 
}
\label{Fig:Dasetal}
\end{center}
\end{figure*}

The east side differs from the west. While Figures \ref{Fig:Halpha} and \ref{Fig:Fe} present a thin shell on the east, a new study by \cite{Dasetal2024Poster} presents new MUSE-VLT images of SNR~0509-67.5 where they resolve two shells in the east, which they claim are compatible with the DDet scenario. Figure \ref{Fig:Dasetal} presents a figure from their work. Panels a and c show the observations in sulfur and calcium lines, while panels b and d are the column density of these elements in the simulation with CO WD mass of $1 M_\odot$ and helium shell mass of $0.03 M_\odot$ from \cite{Collinsetal2022}. \cite{Dasetal2024Poster} claim for similarities between the observations and the simulation, e.g., the sulfur shell on the east (left of panels b and d) is between the two calcium shells in the east (marked ICaS and OCaS on panel c) and therefore, claim for the DDet scenario.  

I find the observations and DDet simulation of \cite{Collinsetal2022} to significantly differ (but note that the simulations are at 100 seconds while SNR~0509-67.5 is hundreds of years old). (1) As the two yellow vertical lines on panels b and d show, in the DDet simulation, the sulfur bright shell has the same inner boundary as the bright calcium shell and a somewhat more extended outer shell. It does not reside between the two calcium shells, as in the observations.    
(2) The DDet simulations show two calcium shells in the bright region (left side in panel d) and one narrow shell on the other side (right side of panel d). The two calcium shells that \cite{Dasetal2024Poster} refer to are in the east (marked ICaS and OCaS on panel c), opposite the bright calcium zone on the west. 

In addition to these challenges of the DDet scenario, there are no indications for a surviving companion in SNR~0509-67.5 (Section \ref{sec:Properties}). The simulations of \cite{Collinsetal2022} are of one exploding WD, and it is unclear if they apply to SNR~0509-67.5. In the case of two exploding WDs (e.g., \citealt{Papishetal2015, Tanikawaetal2019, Pakmoretal2022, Bossetal2024}), at least one in a DDet (which I group here as the DD scenario), explosion is highly non-spherical and not expected to have an elliptical shape (as the ellipses in Figure \ref{Fig:Fe} and \ref{Fig:Dasetal} show for the observed morphology of SNR~0509-67.5).

However, the simulations by \cite{Collinsetal2022} are for limited cases and at 100 seconds after the explosion, long before the development of the reverse shock. The DDet scenario (or the triple detonation channel) is still possible but needs relevant simulations to show its applicability for SNR~0509-67.5.  

\subsubsection{Double shells by instabilities during the explosion}
\label{subsubsec:Instabilities}

I suggest a different possible explanation for the calcium double shell structure. I notice the clumpy structure of the different shells, particularly the outer calcium shell. I also notice the fingers on the east and north, and the columns on the southwest and west that I mark on panel c of Figure \ref{Fig:Fe}. The columns connect the outer Fe arc with the Fe shell. I raise the tentative possibility that Rayleigh-Taylor instabilities formed the calcium double-shell structure during the explosion process. 

\begin{figure}[]
\begin{center}
\includegraphics[trim=0.50cm 17.0cm 5.0cm 0.0cm,scale=0.55]{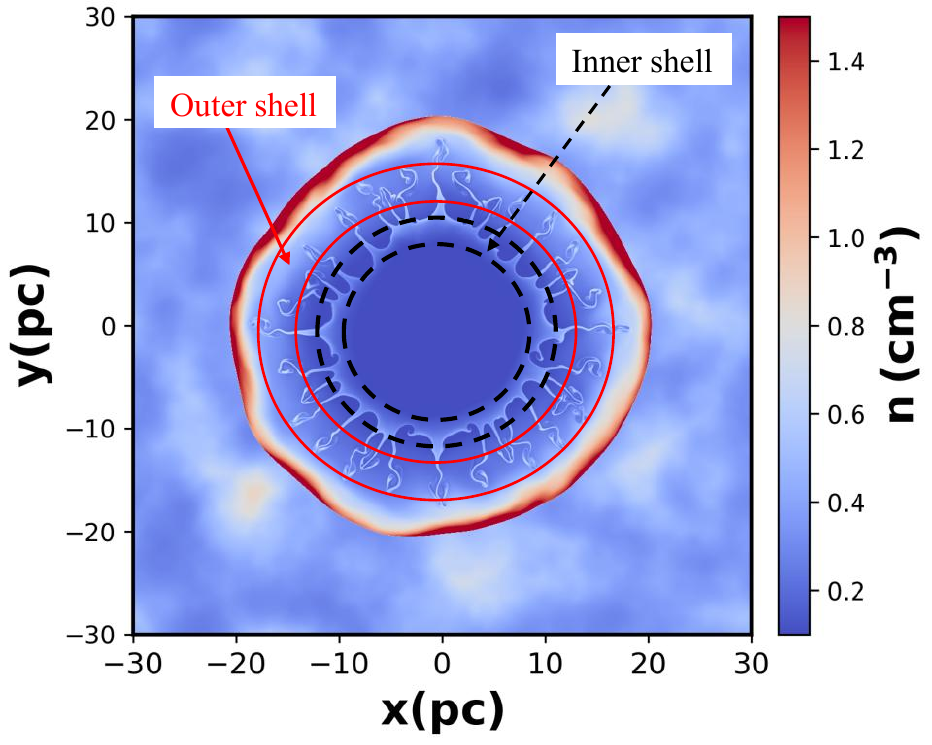} 
\caption{A figure adapted from \cite{Preteetal2025} presenting a density map in a plane through the center of hydrodynamical interaction of supernova ejecta with ambient gas. I mark two rings on the plane that might appear as two shells in observations: An inner high-density zone (white) between the two black-dashed circles that will appear as an inner shell, and the Rayleigh-Taylor instability mushroom's caps that might be observed as an outer shell, mostly between the two red-solid circles. I suggest that such Rayleigh-Taylor instability mushroom caps formed during the explosion process might lead to two calcium-rich shells in SNR~0509-67.5. }
\label{Fig:RTI}
\end{center}
\end{figure}

To encourage simulations of the flow I suggest, I bring as an example a different flow type, which I do not claim is relevant to SNR~0509-67.5, but serves as a demonstration. 
Simulations of supernova ejecta interaction with an ambient gas show many thin Rayleigh-Taylor instability fingers protruding out, then expanding to the side to form a mushroom-type structure (e.g., \citealt{DuffellKasen2017, Preteetal2025}). The caps of the mushrooms form a very clumpy second shell, outer to the main shell. Figure \ref{Fig:RTI}
presents the two shells in a simulation of \cite{Preteetal2025}.  My suggestion is that in the formation of the two shells of SNR~0509-67.5, the Rayleigh-Taylor instability process occurred during the explosion process rather than during the interaction of the ejecta with the ambient gas. The dense inner part is the calcium-rich layer (in the simulation of Figure \ref{Fig:RTI} it is the ejecta), while the lower-density outer region is the sulfur-rich layer of the exploding WD (in the simulation of Figure \ref{Fig:RTI} it is the ambient gas). The lower-density sulfur-rich layer slows down the calcium-rich layer. I suggest that the mushroom caps are the outer calcium shell in SNR~0509-67.5.  
This tentative suggestion requires confirmation with hydrodynamical simulations with an appropriate explosion setting. 

The columns are prominent in the west and the fingers in the east and north. The calcium double-shell appears east, north, and west but not in the south. Nonetheless, the rim in the south is clumpy. I attribute the different appearances in different directions to the different strengths of the interaction with the CSM (the old planetary nebula). It is probably the weakest in the south, hence less energizing the fingers and columns. I speculate that deeper observations will also reveal fingers and/or columns in the south. In addition, the basic nature of instabilities might result in some regions with less developed instabilities during the explosion process. So, another possibility is that the instabilities did not reach the highly non-linear phase in the south.

\cite{Bozzettoetal2014} find a bright inner ring inner to the shells I consider here, which is only seen in the radio continuum, which they find to be non-thermal. Its projection on the plane of the sky is elliptical, with its long axis perpendicular to the long axis of the shells I study here. This inner shell, which might be related to a reverse shock, requires further study.  

\section{Summary}
\label{sec:Summary}

I examined the properties of type Ia SNR~0509-67.5, particularly its morphology (Figure \ref{Fig:Halpha}), to constrain possible progenitor scenarios. 
Based on the similar elliptical morphology of the hydrogen-rich ambient gas and the iron-rich gas inner to it (Figure \ref{Fig:Fe}), I suggested that the ambient gas shaped the SN ejecta and that the explosion was more or less spherical, i.e., had a large-scale deviation from a spherical explosion of $< 4 \%$ (less than half the relative difference between the long and short axis of the ellipse which is $7.5 \%$).  Namely, in SNR~0509-67.5 the CSM shaped the elliptical morphology, not the explosion of the central WD, which I suggest was spherical. 

Based on the elliptical morphology of the hydrogen-rich ambient gas and the higher density near the short axis of the ellipse \citep{Arunachalametal2022}, I raised the possibility that the ambient medium is an old planetary nebula; in that case, the age of the planetary nebula, i.e., the time from the common envelope evolution that ejected the nebula to explosion, is $t_{\rm CEED} \simeq 10^5 - 5 \times 10^5 \yr$ (Section \ref{sec:Properties}). 

Starting with the non-detection of a surviving companion in SNR~0509-67.5 (e.g., \citealt{SchaeferPagnotta2012, Panetal2014, Notaetal2016, Shieldsetal2023}) and the above suggestion that SNR~0509-67.5 is a SNIP (Section \ref{subsec:KeyProperties}),  I discussed the different scenarios from Table \ref{Tab:Table1} as possible progenitors of SNR~0509-67.5 (Section \ref{subsec:Scenarios}). 
My ranking of the possible scenarios is in the last row of Table \ref{Tab:Table1}. The lonely-WD scenarios, namely, the CD and DD-MED scenarios, are the most likely to account for SNR~0509-67.5. Other scenarios encounter challenges (Section \ref{subsec:Scenarios}). 

Despite this ranking from most likely to unlikely, I emphasize that I discussed all scenarios as I cannot completely rule them out. I reiterate my call to consider all these scenarios when analyzing specific SNe Ia.  

I suggested that the double-shell structure that \cite{Dasetal2024Poster} observed in calcium (Figure \ref{Fig:Dasetal}) results from Rayleigh-Taylor instability mushroom caps formed during the explosion process; The caps of the mushrooms form the outer shell (Figure \ref{Fig:RTI}; Section \ref{subsubsec:Instabilities}).

The suggestion of SNR~0509-67.5 being an SNIP, if it holds, has further implications, as follows.

(1) The fraction of SNIPs might be higher than the previously estimated 50 percent of all normal SNe Ia \citep{Soker2022RAA}. I refer here to normal SNe Ia, not to peculiar SNe Ia.  The suggested larger fraction of SNIPs is supported by the larger estimated number of SNe Ia interacting late with a CSM (e.g., \citealt{Moetal2025}). 
  
(2) The higher fraction of SNIPs, $f_{\rm SNIP} > 0.5$, together with the suggestion that SNR~0509-67.5 comes from a younger population, $\tau_{\rm SF} \lesssim 5 \times 10^8 \yr -  1.5 \times 10^9 \yr$ (Section \ref{subsubsec:CDDDMED}), than the general population in its larger environment $\tau_{\rm SF} \simeq 7.9 \times 10^9 \yr$ \citep{Badenesetal2009}, implies that some pockets of more recent star formation exist in old populations, possibly including elliptical galaxies. This is compatible with cooling flows in elliptical and clusters of galaxies (e.g., \citealt{Reefeetal2025}).  This point is significant as it implies that SN Ia scenarios where the secondary star is also massive, like the CD scenario, might be applicable even in galaxies with general old stellar population, like elliptical galaxies.  This speculation deserves further study. 

(3) Instabilities,  mainly Rayleigh-Taylor instabilities,   during the explosion process might form clumps of some elements moving much faster than the mean velocity of these elements. Such clumps might explain fast, intermediate mass elements that \cite{Hoogendametal2025} argue to be common in SNe Ia, based on their observations of calcium and silicon features near velocities of $\simeq 0.1 c$ in SN Ia 2024epr (see also \citealt{Iskandaretal2025}).

\section*{Acknowledgments}
I thank Ashley Ruiter for useful comments,  and an anonymous referee for very useful comments and suggestions that improved the presentation of the results.  A grant from the Asher Space Research Institute in the Technion supported this study. 





\end{document}